\documentstyle[12pt,psfig,aasms4]{article}
\def\apgt{\ {\raise-.5ex\hbox{$\buildrel>\over\sim$}}\ }
\def\aplt{\ {\raise-.5ex\hbox{$\buildrel<\over\sim$}}\ }

\newcommand{\mc}{\mbox {$M_{Ch}$}}
\newcommand{\pyr}{\mbox {{\rm yr$^{-1}$}}}
\newcommand{\md}{\mbox {$\dot{M}$}}
\newcommand{\myr}{\mbox {~${\rm M_{\odot}~yr^{-1}}$}}
\newcommand{\ace}{\mbox {$\alpha_{ce}$}}
\newcommand{\ms}{\mbox {M$_{\odot}$}}
\newcommand{\msun}{\mbox{${\rm M}_\odot$}}

\begin{document}
\title{Supernovae Rates: A Cosmic History}

\author{Lev R.\ Yungelson$^1$ and Mario Livio}
\affil{Space Telescope Science Institute\\
3700 San Martin Drive\\
Baltimore, MD 21218}
\altaffiltext{1}{Permanent address: Institute of Astronomy of the Russian Academy of
Sciences, 48~Pyatnitskaya Street, 10917 Moscow, Russia.}

\begin{abstract}
We discuss the cosmic history of supernovae on the basis of various 
assumptions and recent data on the star formation history.

We show that supernova rates as a function of redshift can be used to 
place significant constraints on progenitor models, on the star formation 
history, and on the importance of dust obscuration.  

We demonstrate that it is unlikely that the current observational indications 
for the existence of a cosmological constant are merely an artifact of 
the dominance of different progenitor classes at different redshift intervals.  
\end{abstract}

{\it Subject headings:} binaries: 	close -- 
	               stars: 	        formation  --  
	               supernovae:      general --
                           cosmology: observations

\section {Introduction}

The interest in the cosmic history of supernovae stems from several sources. 
First, core-collapse supernovae (types~II and Ib/c) directly follow the 
star formation history and some of them may be related to gamma-ray bursts. 
Secondly, Type~Ia supernovae (SNe~Ia) are being used as the primary standard 
candle sources for the determination of the cosmological parameters $\Omega$ 
and $\Lambda$ (e.g.\ Perlmutter et~al.\ 1999; Riess et~al.\ 1998). Thirdly, 
a comparison of SNe~Ia rates (for the different models of their progenitors) 
with observations may shed light on both the star formation history and on 
the  nature of the progenitors (e.g.\  Yungelson \& Livio 1998; Madau 1998a). 
Finally, the counts of distant SNe could be used to constrain cosmological 
parameters (e.g.\  Ruiz-Lapuente \& Canal 1998). As a consequence of the above, 
studies of cosmological SNe are among the primary targets for the Next Generation 
Space Telescope (NGST), which presumably will be able to detect, with proper filters,  
virtually all the SNe up to a redshift $z\sim 8$\ (see e.g.\ http://ngst.gsfc.nasa.gov/Images/sn.GIF).

In the present study we combine data on the precursors of SNe~Ia in our Galaxy 
with data on the cosmic star formation rate in an attempt to analyse the 
frequency of events as a function of redshift.  

In view of the uncertainties that still exist concerning the cosmic star 
formation history, we use two types of inputs to characterize the star 
formation rate (SFR). In the first, we use profiles inferred from deep 
observations (e.g.\ Madau, Panagia \& Della Valle 1998). In the second, 
we use a step-wise SFR which includes a burst of star formation and a 
subsequent stage of a lower SFR. In the latter case the star formation 
history is parameterized by the duration of the star  burst  phase and 
by the fraction of the total mass of the stellar population that is formed 
in the burst.

The different scenarios for SNe~Ia are briefly discussed in \S2. The basic assumptions 
and model computations are presented in \S3 and \S4, and a discussion and conclusions follow.

\section {The Progenitors of SNe Ia}

The observed SNe~Ia do represent somewhat of a mixture of events, with a majority of 
``normal'' ones and a small minority of ``peculiar'' ones (see e.g.\ Branch 1998, and 
references therein). A more moderate  diversity is present even among the ``normals.'' 
There exist certain relations between the absolute magnitudes and light curve decline rates 
and the morphological types of  the host galaxies (e.g.\ Branch, Romanishin, \& Baron 1996; 
Hamuy et~al.\ 1996). This may suggest a possible diversity among the progenitors of SNe~Ia 
(see e.g.\ review by Livio 1999).    

On the theoretical side, SNe Ia are very probably thermonuclear disruptions of accreting 
white dwarfs. Two classes of explosive events are generally considered in the literature.   
The first involves central ignition of carbon when the accreting white dwarf reaches the 
Chandrasekhar mass $\mc \approx 1.4~\msun.$ In the second, the ignition of the accreted 
helium layer on top of the white dwarf induces a compression of the core which leads to 
the ignition of carbon at sub-Chandrasekhar masses (these are known as edge-lit detonations 
or ELDs). Parameterized models for the events in the former class are able to reproduce 
most of the typical features of SNe~Ia, while ELD models encounter a few serious problems 
when confronted with observations (see e.g.\ H\"{o}flich \& Khokhlov 1996; Nugent et~al.\ 1997; 
Branch 1998; Livio 1999 for a discussion and references). On the other hand, binary evolution 
theory clearly predicts situations in which helium may accumulate on top of white dwarfs 
(see e.g.\  Branch et~al.\ 1995; Yungelson \& Tutukov 1997). It is presently not entirely 
clear whether ELDS indeed do not occur in nature, or whether they are responsible for a 
subset of the events (e.g.\ the subluminous ones). However, both the existing diversity 
in the observed properties of SNe~Ia and the uncertainties still involved in theoretical 
models, suggest that it is worthwhile to explore all the possible options.  

The occurence rate of SNe~Ia inferred for our Galaxy is $\sim 10^{-3}$ \pyr\ (Cappellaro 
et~al.\ 1997). There are three evolutionary channels in which according to population 
synthesis calculations the realization frequency of potentially explosive configurations 
in the disk of the Milky Way is at least at the level of $10^{-4}\,\pyr.$  These are:

{\bf A}. Mergers of double degenerates resulting in the formation of a $M \apgt\mc$\ object 
and central C ignition.    The channel involves the accretion of carbon-oxygen.

{\bf B}. Accretion of helium from a nondegenerate helium-rich companion at a rate of 
$\md \sim 10^{-8} \myr$, resulting in the accumulation of a He layer of $\sim (0.10 - 0.15)\,
\msun$ and an ELD.

{\bf C}. Accretion of hydrogen from a (semidetached ) main-sequence or evolved companion. 
The burning of H may result either in the accumulation of \mc\ and central C ignition or 
in the accumulation of a critical layer of He for an ELD.

The positive aspects and draw-backs of these channels were discussed in detail elsewhere 
and also by other authors (e.g.\ Tutukov, Iben, \& Yungelson 1992; Branch et~al.\ 1995; 
Iben 1997; Ruiz-Lapuente et~al.\ 1997;  Yungelson \& Livio 1998; Hachisu, Kato \& Nomoto 
1999; Livio 1999). Here we present for ``pedagogical'' purposes a simplified flow-chart 
which illustrates some of the evolutionary scenarios which may result in  SNe~Ia (Fig.~1). 
Other channels may definitely contribute to the total SNe~Ia rate but they are either 
less productive or they involve large uncertainties (see also \S5).

In a typical scenario, one starts with a main-sequence binary in which the mass of the 
primary component is in the range $\sim (4 - 10) $\,\msun.  The initial system has to be 
wide enough to allow the primary to become an Asymptotic Giant Branch (AGB) star with a 
degenerate CO core. After the AGB star overfills its Roche lobe a common envelope forms. 
If the components do not merge inside the common envelope, the core of the former primary 
becomes a CO white dwarf. The subsequent evolution depends on the separation of the components 
and on the mass of the secondary.  If the latter is higher than $\sim 4$\,\msun\ and the 
secondary fills its Roche lobe  in the AGB stage, then following a second common envelope 
phase, a pair of CO white dwarfs forms. The two white dwarfs may merge due to systemic 
angular momentum losses via gravitational wave radiation. As a result, a Chandrasekhar 
mass may be accumulated, leading potentially to a SN~Ia (scenario~{\bf A}). 

If the  mass of the secondary is above $\sim 2.5$\,\msun\ and it fills its Roche lobe before  
core He ignition it becomes a compact He star. If inside the common envelope the components 
get sufficiently close, the He star may fill its Roche lobe in the core He burning stage 
and transfer matter at a rate of $\sim 10^{-8}$\,\myr. The accumulation of He on top of 
the white dwarf may result in an ELD (scenario~{\bf B}).   

Finally, if the mass of the companion to the white dwarf is below $\sim (2 - 3)$\,\msun, 
the companion may fill its Roche lobe on the main-sequence or in the subgiant phase. Such 
a star could stably transfer matter at a rate which allows for the accumulation of \mc, 
or of a critical-mass He layer (scenario~{\bf C}). 

Below we refer to all the potentially explosive situations listed above as ``SNe~Ia,'' 
in spite of the fact that it is not entirely clear whether most of these configurations 
actually result in a SN (see e.g.\ Livio 1999). We should note that while the above quoted 
masses are only approximate, the uncertainties are not such that they can change the 
expected rates significantly.

A special remark has to be made concerning the exclusion of symbiotic stars. Yungelson 
et~al.\ (1995) have shown that the accumulation of \mc\ in these systems occurs at a low 
rate: $\sim 10^{-5}$\,\pyr\ (see however discussion in \S5). The accumulation of 0.15~\msun\ 
of He via H burning occurs at a rate of $\sim 
10^{-4}$\,\pyr, but the accretion rate is typically high, and hence, one would normally not 
expect an ELD to ensue. Rather, weak helium flashes may occur. A cautionary note has also 
to be made concerning ELDs which under certain sets of parameters have an occurence rate 
of $\sim 10^{-3}$\,\pyr\  in semidetached systems (scenario~{\bf C}, Yungelson \& Livio 1998). 
We assumed that ELDs occur even if the accretion rate of hydrogen was initially high but 
then dropped to below $3 \times 10^{-8} \myr$. By this, we neglected the possible influence 
of hydrogen flashes on the helium layer. The response of the helium layer and the underlying 
white dwarf  to the  varying accretion rate (from several times  $10^{-7}$ \myr\ to 
$3\times 10^{-8}$ \myr) was never treated in detail to the best of our knowledge. One may 
expect a competition between cooling (due to the expansion of the hydrogen layer) and the 
inward heat propagation (due to nuclear burning).  

Cassisi, Iben, \& Tornamb{\`e} (1998) for example, claim that heating by hydrogen flashes 
keeps the temperature of the He layer high and may even prevent the explosive ignition of 
He.  Rather, they conclude, quiescent burning may be expected (for accretion rates $10^{-8}$
--$10^{-6}$~M$_{\odot}$~yr$^{-1}$) during which the white dwarf expands to giant dimensions 
and its envelope may be removed by interaction with the companion. If an explosion nevertheless 
happens, it may produce a powerful nova-type event (a ``super nova''). As a result of all 
of these uncertainties (and others) the issue of ELDs via a channel of hydrogen accretion 
is not definitively settled (see Livio 1999). 

One of the cornerstones of channel~{\bf C}  is the assumption of negligible mass loss in 
the form of a wind during helium flashes (e.g.\ Kato, Saio, \& Hachisu 1989), which allows 
for the accumulation of \mc\ despite the flashes. The expansion of the helium layers found 
by Cassisi et~al.\ and the accompanying mass loss may (in some cases at least) prevent the 
accumulation of \mc.

Thus, the realization frequency of both scenarios for SNe~Ia (explosion at \mc\ or at a 
sub \mc\ mass) via channel~{\bf C} is a matter of considerable uncertainty. Nevertheless, 
we include channel~{\bf C} in our consideration (although see discussion in \S5). 
 
The basic difference between the possible progenitor scenarios of SNe Ia is in the ``evolutionary 
clock''---the time interval between the formation of the binary system and the SN explosion.  
Figure~2 shows the dependence of the supernova rate on time after an instantaneous star 
formation burst, for the four mechanisms listed above, as computed in the present study. 
The curves shown were computed for a common envelope efficiency parameter $\ace = 1$; the 
dependence on this parameter within reasonable limits on \ace\ between 0.5 and 2 is not too 
strong.  For semidetached systems, we considered the case of mass exchange stabilized by 
the presence of a thick stellar wind (Hachisu, Kato, \& Nomoto 1996, henceforth, HKN) as 
modified by Yungelson and Livio (1998). Further suggested modifications to the standard 
evolution will be discussed in \S5.  The differences in the timespan between the formation 
of a binary and the SN~Ia event, and in the rate of decay of the SNe rates in the different 
channels, manifests itself in the redshift dependence of the SNe~Ia rates.

Our calculations are based on the assumption that the IMF, and the mechanisms of SNe~Ia  are 
the same throughout the Hubble time. This assumption may not be valid, for example because of 
metallicity effects. Stars with lower $Z$ develop larger helium and  carbon-oxygen cores for
the same main-sequence mass (e.g.\ Umeda et~al.\ 1998), and hence, form more massive white 
dwarfs. At the same time, the upper mass limit of stars which form CO white dwarfs decreases 
towards a lower metallicity. However, assuming a power-law IMF both of these effects result 
in an increased number of potential pre-SNe~Ia white dwarfs. On the other hand, a low metallicity 
can inhibit strong optically thick stellar winds which are essential for the HKN model of SNe~Ia 
(Kobayashi et~al.\ 1998). Assuming for the moment that several channels may contribute (see 
however Livio 1999), the net effect may be an enhanced rate of SNe~Ia from the channels of 
double-degenerates and ELDs from systems with nondegenerate He donors, and a reduction in 
the rate from the channel of hydrogen-donor systems. 

\section{SNe and the Star Formation rate}
\vspace{-6pt}
\subsection{Supernovae Rates}
The rest-frame frequency of SNe of a certain type  at any time $t$, $n(t)$, may be derived 
by convolving the star formation rate $\Psi(\tau)$ with the function $f(t)$ giving the rate 
of SNe after an instantaneous burst of star formation: 
\begin{equation} 
\vspace{-16pt}
n(t)= \int_0^t f(t-\tau) {\rm \Psi(\tau) } d\tau.  
\vspace{-4pt}
\end{equation}

Two approaches for the evaluation of $f(t)$\ are encountered in the literature. The first 
is not to consider any specific mechanisms of SNe (which are still a matter of some debate),  
but rather to parameterize $f$ by the fraction of exploding stars in the binary star population 
(the ``explosion efficiency'')  and the delay between formation and explosion, or the 
``evolutionary clock'' (e.g.\ Madau et~al.\ 1998; Dahl{\'e}n \& Fransson 1998;  Sadat et~al.\ 1998). 

For core-collapse supernovae (SN~II and SN~Ib/c) it is natural to assume that the shape of 
$f$\ follows the SFR and the delay between the formation of the star and the SN event is 
negligible, since the lifetime of stars more massive than 10~\ms\ is $\aplt 20$Myr.

For SNe~Ia Madau et~al.\ considered a parameterized $f(t)$ with timescales of 0.3, 1, and 
3~Gyr between the formation of the WD and the explosion. These authors reproduce the ratio  
of ${\rm SN_{II}/SN_{Ia} \approx 3.5}$\ in the local Universe, if the explosion efficiency 
is 5\% to 10\%.

A similar parameterization was adopted by Dahl{\'e}n \& Fransson (1998), who estimated the 
number of core-collapses and type Ia~SNe which may be detected by NGST in different filters 
for different limiting stellar magnitudes. 

Sadat et~al.\ (1998) considered a power-law $f \propto t^{-s}$, and explored a range of $s$\ 
from 1.4 to 1.8.   Another parameter of Sadat et~al.\ is the rise time of the SNe Ia rate 
from 0 to a maximum which was fixed at 0.75~Gyr. The ranges of $s$ and rise times were 
derived from models of the chemical evolution of Fe in elliptical galaxies and in clusters 
of galaxies. Concerning the explosion efficiency, Sadat et~al.\ actually do not exploit 
this parameter since they additionally normalize their rates, in order to reproduce the 
local rate of SNe~Ia by the adopted SFR. 

A different approach to the determination of $f$ relies on population synthesis calculations. 
Using this method, J{\o}rgensen et~al.\ (1997) derived the rates of core-collapse supernovae 
(SNe~II and SNI~b), mergers of binary WDs with a total mass exceeding \mc, and collapses of 
Chandrasekhar mass white dwarfs in semidetached systems (in the standard model, without the 
thick wind of HKN).  

Ruiz-Lapuente \& Canal (1998; see also Ruiz-Lapuente, Canal, \& Burkert 1995; Canal, 
Ruiz-Lapuente, \& Burkert 1996) considered as SNe~Ia progenitors merging double degenerates 
and cataclysmic binaries. For the latter channel, $n(t)$ was estimated in two cases. First,  
the ``standard'' case which allows only  thermally stable mass exchange for $M_{\rm donor}/M_{\rm accretor} 
\aplt\break 0.78.$ Second, the case of the ``wind'' solution of HKN, which allows for mass 
exchange at rates of up to $\sim 10^{-4} \myr,$ for systems with $q \aplt 1.15.$  Ruiz-Lapuente 
\& Canal find a distinct difference between the behavior of the predicted SN~Ia rates {\it vs\/}. 
limiting red stellar magnitude for different families of progenitors. Namely, the $dN/dm_R-m_R$ 
relation for descendants of cataclysmic variables is much steeper than that for merging 
double degenerates. However, the computations of Dahl{\'e}n \& Fransson (1998) do not show 
any significant difference in the behavior of the SNe~Ia counts for different delays in the 
0.3--3~Gyr range (the main difference between double degenerates and cataclysmic variable 
like systems is in the delay time).   

\subsection{The star formation rate}

The star formation rate which is used as an ingredient in calculations of the evolution of 
the cosmic SNe rate is usually derived from studies which model  the observed evolution of 
the galaxy luminosity density with cosmic time. For example, Madau, Pozetti, \& Dickinson 
(1998) and Madau, Della Valle, \& Panagia (1998, hereafter MDVP98) have shown that the  
observational data can be fitted if one assumes, as an ingredient of the model, a time-dependent 
star formation rate. However, there exist uncertainties in this model, due to the uncertain 
amount of dust extinction at early epochs. For example, MDVP98 have shown that the same 
observational data may be fitted if one assumes a constant $E_{\rm B-V} = 0.1$ or a $z$-dependent 
dust extinction which rises rapidly with redshift, $E_{\rm B-V} = 0.011(1+z)^{2.2}$. The 
latter authors provide convenient fitting formulae for the star formation rates for these two cases.

Model 1 (``little dust extinction'') has
\begin{equation}
\Psi(t)=0.049 t^5_9 e^{-t_9/0.64}+0.2(1 - e^{-t_9/0.64})\ {\rm \msun\,yr^{-1}\, Mpc^{-3}}~~,
\end{equation}
where $t_9$ is the time in Gyr, $t_9 = 13(1+z)^{-3/2}$.  
 
Model 2 (``$z$-dependent dust opacity'') has
\begin{equation} \Psi(t)=0.336 e^{-t_9/1.6}+0.0074(1 -
e^{-t_9/0.64})+0.0197t^5_9 e^{-t_9/0.64}{\rm \msun\,yr^{-1}\,Mpc^{-3}}~~.
\end{equation}

Note that eqs.~(2) and (3) give slightly different current SFR and the integrated values 
are also different by about 10\%. Both SFR models predict a similar, rather steep rise, by 
a factor $\sim 10$,  at $z \aplt 1.5$. The difference between the two rates is in the behavior 
at $z \apgt 1.5$. While in the ``little dust extinction'' case the rate drops almost linearly 
by a factor of about 10 to $z_\star = 5$, in the ``$z$-dependent dust opacity'' case it 
continuously grows to $z_\star$, by a factor of $\sim 2.5$. Formally, the star formation 
process switches on discontinuously at $z_\star$.

We should note that  Model~2 may be a more realistic representation of the global star 
formation history, since there is growing evidence of a significant effect of dust absorption 
at high~$z$ (e.g.\  Pettini et~al.\ 1998; Calzetti \& Heckman 1998; Huges et~al.\ 1998; Steidel 
et~al.\ 1998; Blain et~al.\ 1998). Also, selection effects due to the low surface brightness 
of galaxies (e.g.\ Ferguson 1998) or the shift of typical spectral features  to the red 
(e.g.\ Hu, Cowie, \& McMahon 1998) may result in an underestimate of the SFR at high redshifts. 

Equation (2) gives a star formation history which is consistent with expectations from 
hierarchical clustering cosmologies, while Eq.~(3) gives the model prediction for SFR 
typical for a monolitic collapse scenario (e.g.\ Madau 1998b).  

Ruiz-Lapuente \& Canal (1998) used in their computations the star formation rate given by 
Madau (1997), without corrections for dust extinction. The effect of extinction was considered 
by Dahl{\'e}n \& Fransson (1998) and by Sadat et~al.\ (1998). In the latter case, the SFR 
at $z \apgt 1$ was taken to be several times higher than in the ``low-dust'' case. 
J{\o}rgensen et~al.\ (1997) considered two modes of star formation: a ``burst'' lasting 
for 500~Myr, and a continuous SFR for a Hubble time, and computed models for a range of relative contributions of both star 
formation modes. 

\section {Model computations}
\subsection{Models using ``observed'' star formation rates}
The population synthesis code used for the computations of the SNe rates was previously 
applied by the authors to a number of problems related to the population of galactic binary 
stars and, in particular, to SNe.  Within the range of observational uncertainties, the 
code reproduces correctly the rates of SNe inferred for our galaxy (Tutukov, Yungelson, 
\& Iben 1992; Yungelson \& Livio 1998 and references therein).

Throughout this paper we assume a cosmology with $\Omega_0=1$, $H_0=50\,{\rm km\, s^{-1} 
Mpc^{-1}}$. These values of the cosmological parameters are assumed only for convenience. 
Our qualitative results and conclusions do not depend on this choice. Star formation is 
assumed to start at $z_\star =5$. The Hubble time in this model was taken to be 13~Gyr.

For the different SNe~Ia scenarios listed in \S2 and for different star formation histories, 
we first calculated the rest-frame rates of events $n_0$. We then computed differential 
functions for the number of events observed at redshift $z$ and cumulative functions  
$n(< z)$. We use  eq.~[3.3.25] from Zel'dovich \& Novikov (1983) for the number of events 
observed from a layer between redshifts $z$ and $z+dz$ in an expanding, curved Universe, 
taking into account time dilation:

\begin{equation}
\frac {dn} {dz} = n_0 \frac{4 \pi c^3} {H_0^3} \frac {1}{1+z}
\xi_z (z,\Omega_0) z^2 dz~~.
\end{equation}
Where, for the particular case of $\Omega_0=1$
\begin{equation} 
\xi_z(z,1)= \frac{4[(1+z)^{1/2}-1]^2}
{z^2(1+z)^{5/2}}~~.
\end{equation}
Notice, that $\frac {\partial \xi_z} {\partial z}  <0$.  
Time $t$ is related to $z$ as
\begin{equation}
t = \frac{2 H_0^{-1}}{3(1+z)^{3/2}}~~.
\end{equation} 

We operate with the {\it number of events per yr\/} instead of expressing the SNe rates in 
the more conventional Supernovae Units (SNU), since both the  computation of  blue luminosities
and their observational determination involve additional parameters (expressing the rates 
in SNU may result in loss of information on both the SNe rates and on the SFR).  

Our simulations give the rates of SNe as a function of $z$. Clearly, the number of observable 
events depends on other factors such as the limiting stellar magnitude of the sample, etc. 
Nevertheless, our results provide the basis for theoretical expectations, which need subsequently 
to be convolved with observational selection effects. In principle, NGST observations can approach 
the theoretical limits. Figure~3 compares the values of $dn/dz$ for the different channels 
of SNe and the different assumptions about the SFR given by eqs.~(2) and~(3). Figure~4 shows 
the behavior with redshift of the cumulative numbers of SNe.  

The behavior of $dn/dz$ can be understood as follows. In Model~1 (low dust) as one progresses 
from $z=0$ to $z_\star$,  the SFR reaches a maximum at $z \approx 1.5.$ The maxima of the 
rest-frame SNe rates happen at a slightly lower $z$ in order of decreasing delay times: 
ELDs in systems with subgiant companions, \mc\ accumulations in the latter,  mergers of 
double degenerates, ELDs in systems with nondegenerate He donors, core-collapse SNe\  
(Fig.~2). The behavior of the $dn/dz$ counts depends also on the  geometrical 
$z$-dependent factors given by eqs.~(4) and (5). In particular, the derivative 
of the product $z^2\xi_z$ changes sign from positive to negative at  $z \approx 0.96$. 
This factor shifts the maximum in the counts to a lower $z$. The steep rise of $dn/dz$ at 
low $z$ is entirely due to the expanding horizon.

Similarly, in Model~2 ($z$-dependent dust opacity),  the behavior of $dn/dz$ at low $z$ 
is dominated by the expansion of the comoving volume and the rates suggested by the two 
models are almost indistinguishable. However, already at $z \sim 0.5$, the increase in the 
rates in Model~2 becomes somewhat less steep, reflecting the more moderate growth of the SFR. 
The rate of core-collapse SNe starts to decrease at $z \approx  1.2$\ despite the continuous 
growth of the SFR. This is a consequence of the negative $\partial\xi_z\over \partial z$. 
The rates of SNe~Ia start to decline at a higher $z$, as a consequence of the longer delay 
times.  The difference in the time delays between SNe~II and the different hypothetical SNe~Ia 
manifests itself in an increase in the SN~Ia/SN~II ratio at low $z$\ and its subsequent 
decline (Fig.~7). This feature was already noticed by Yungelson \& Livio (1998) for SNe~Ia 
from double degenerates, but in the present study we find that (i)~this effect is less 
pronounced due to the different approximation to the SFR and (ii) the redshift of the 
maximum of the ratio is different for different SNe~Ia scenarios. 

The difference in the rate of decline of  $dn/dz$  at $ z \apgt 1$ is clearly distinct in 
Models~1 and~2 and may provide  important information about the star formation behavior.

The most pronounced feature of $dn/dz$  for both types of dust models is the disappearance 
of SNe~Ia at $z \approx 3$ for the channels of progenitors with relatively long delays. 
Thus, in principle, a determination of SNe~Ia  rates at $z\apgt3$ with NGST can unambiguously 
distinguish between different progenitor models. Long delay times are typical for both modes 
(Chandrasekhar or sub-Chandrasekhar explosions) of SNe~Ia resulting from systems with subgiant 
donors. 

The relative role of different channels for SNe Ia changes with $z$. In both models 1 and 
2 (for the dust) mergers of double degenerates dominate over ELDs in systems with He nondegenerate 
donors up to  $z \aplt 0.4$. In Model~1, ELDs in systems with subgiants dominate over He-ELD 
at $z \aplt 0.8$ and over DD-Ch at $z\aplt 1.3$. In Model~2 these limits are  at about 
$z\sim1$ and $z\sim2.2$. If it were the case that all three channels really contribute to 
SNe~Ia but have somewhat different characteristics, one would expect to find variations in 
the statistical properties of SNe~Ia with redshift. We will return to this point in the discussion.

As expected, the cumulative numbers of SNe grow faster in the ``low dust'' case than in the 
model with dust. Only the cumulative counts of SNe~II and SNe~Ia from the DD-Ch  and He-ELD 
channels grow continuously to high redshifts, while those for SNe involving subgiants 
saturate at $z \sim 3$. 

To summarize this section: observations of  SNe beyond $z \approx 1$ can provide valuable 
information on the star formation rate (see also \S5). The counts of SNe~Ia at $z \apgt 3$
will indicate the timescale of the delay between births of  binaries and SN events and will 
then provide information on the nature of the progenitors.  

\subsection{Parameterized star formation rates}
The main uncertainty in the global SFR is due to the effects of dust obscuration in star-forming 
galaxies (see e.g.\  Calzetti \& Heckman 1998; Pettini et~al.\ 1998; for  a discussion of 
the fraction of light absorbed by dust). Therefore, it is worthwhile to investigate the 
cosmic history of SNe for several parameterized SFR. 
 
We consider four parameterized modes of galactic star formation (intended to bracket and 
cover a range of possibilities):

  Model 3---constant star formation rate from $z_{\star}=5$ to $z=0$;

  Model 4---a star formation burst which begins at $z_\star$ and has a constant SFR for 1~Gyr;

  Model 5---a star formation burst which begins at $z_\star$ and has a constant SFR for 4~Gyr;

  Model 6---an initial star formation burst which lasts for 4~Gyr with a constant SFR and 
converts 50\% of the total mass into stars, followed by another stage of a lower constant 
SFR which produces the remaining 50\% of the stars (``step-wise SFR'').

For all the cases we normalize the SFR in such a way that the total amount of matter 
converted into stars is equal to the integral over time of Eq.~(2). The overall normalization 
is of no real significance however, since we are interested in the qualitative behavior of SNe counts.

The computations provide us with information on the behavior of SNe rates with redshift, 
for different star formation histories. The results provide insights into the understanding 
the SNe histories in galaxies of different morphological types, which show a wide variety 
of star formation patterns both along the Hubble sequence and within particular classes 
(e.g.\  Sandage 1986; Hodge 1989; Kennicutt, Tamblyn, \& Congdon 1994; Kennicutt 1998). 
Even  among the Local Group dwarf galaxies one encounters very different star formation 
histories, including early bursts, almost constant SFR, and step-wise ones (e.g.\ Mateo 1998).

Figures 5 and 6 present the number counts of SNe Ia per unit $\Delta z$ and the cumulative 
rates of events $n(<z)$ for the above models. The results can be summarized as follows. 

1. Models 3 and 4, with initial starbursts of different durations $\Delta \tau$, clearly 
predict an abrupt decline in the SNe~II rate when moving from $z_{\star}$\ to lower redshifts, 
reflecting the cessation of star formation. The redshift of this sharp decline in the rate 
indicates (for given cosmological parameters) the value of $\Delta \tau$. It also depends 
of course on $z_{\star}$. The behavior of SNe~Ia from ELDs in systems with He donors (He-ELD) 
and Chandrasekhar mass SNe in systems with subgiant donors (SG-CH) shows a similar decline, 
but shifted to a lower $z$ and less abrupt.   

For stellar populations with strong initial star formation bursts this means that, if He-ELD 
and SG-Ch SNe~Ia were the only mechanisms for SNe~Ia, then as one advances to higher redshifts, 
first the rate of SNe~Ia and then the rate of SNe~II would rapidly rise. In the case of SG-Ch 
SNe~Ia the rate would rapidly decline at $z \sim 3.3$. Such a behavior of the rates of SNe~Ia 
in E-S0 galaxies would provide evidence supporting the SG-Ch mechanism for SNe~Ia. 

2. Strong initial peaks in the SFR followed by a variation of the SFR on a short timescale 
manifest themselves in changes in $d^2n/dz^2$ for SNe~Ia. These changes are delayed (to 
lower redshifts) compared to the decrease in the SNe~II rates.

3. SNe~Ia from the mergers of double degenerates (DD-Ch) are the only types of events which
may show up close to $z_{\star}$ and which continue to $z=0$ irrespective of the star formation mode.

4. Although SNe~Ia from ELDs in subgiant systems (SG-ELD) start to explode only at $z\approx 3$, 
in the constant SFR and step-wise SFR models, at $z \aplt 2$ the distribution of their number 
counts {\it vs\/}.\ $z$ becomes very similar to the one from double degenerates (DD-Ch) both 
in morphology and in amplitude. These SNe, however, follow the variations in the SFR slightly 
slower. These two types of SNe~Ia are the only events which may be present at low $z$ even 
if the star formation process ceased long ago (see however \S5).

5. SNe Ia from collapses of Chandrasekhar mass white dwarfs in subgiant systems \mbox{(SG-Ch)} 
may be observed only if the star formation still continues or ceased less than $\simeq 2$~Gyr 
ago (see however \S5).

The same is true for SNe from ELDs in systems with nondegenerate donors \mbox{(He-ELD)}. 
A fast decline of the SNe~Ia rate shortly after (at lower $z$) the decline of the SNe~II 
rate would indicate that either He-ELD or SG-Ch occur.       

6. In the case of a SFR which was almost constant during the past several Gyr there is no 
decline in the SNe~Ia/SNe~II ratio between $z=0$ and 1.

7.  The difference between the behavior of $dn/dz$ for a constant-rate and for step-wise 
SFR models is not significant. This means that only a very significant increase in the 
SFR towards high~$z$\ (like in Model~2) may be reflected in the behavior of the differential 
SNe counts. On the other hand, a rapid decline in the SFR beyond a certain redshift (like 
in Model~2) can be detected easily.

8.  The counts of $dn/dz$ in the redshift range $z \aplt 0.2$--0.4 can hardly provide any 
information about the SFR since they are dominated by the increase of the comoving volume.

\section{Discussion and Conclusions}

Since observations of SNe~Ia are now being used as one of the main methods for the determination 
of cosmological parameters, the importance of identifying the progenitors of SNe~Ia cannot 
be overemphasized. We have shown that different progenitor models result in different SNe~Ia 
rates (or different ratios of frequencies of SNe~Ia to those resulting from massive stars) 
as a function of redshift. One key difference, for example, is in the fact that in all the 
models that involve relatively long delays between the formation of the system and the SN 
event (e.g.\ models with subgiant donors), the ratio $R(SNe~Ia)/R(SNe~II, Ia, Ic)$ decreases 
essentially to zero at $z\apgt3$ (Fig.~7). Thus, future observations with NGST will in principle 
be able to determine the viability of such progenitor models on the basis of the frequencies 
of SNe~Ia at high redshifts.

Probably the most important question that needs to be answered is the following: assuming 
that two (or more) different classes of progenitors may produce SNe~Ia, is it possible that 
the rate of SNe~Ia is entirely dominated by one class at low redshifts ($z<0.5$) and by 
another at higher redshifts ($0.5\aplt z\aplt1.2$)? Clearly, if this were the case, then 
the suggestion of a cosmological constant would have to be re-examined (SNe~Ia at the higher 
$z$ only need to be systematically dimmer by $\sim0.25$~mag to mimic the existence of a 
cosmological constant). An examination of the qualitative behavior of the rates in Fig.~7 
reveals that {\it in principle\/}, the rate at low redshifts could be dominated by ELDs, 
while the rate at higher redshifts by coalescing double-degenerates. However, if ELDs produce 
SNe~Ia at all, these are probably of the underluminous variety (like SN~1991bg; e.g.\ Nugent 
et~al.\ 1997; Livio 1999; Ruiz-Lapuente et~al.\ 1997). Therefore, a division of this type 
would produce exactly the opposite effect to the one required to explain away the need for 
a cosmological constant (the high redshift ones would be brighter). A second possibility 
is that the rate of SNe~Ia resulting from the accumulation of \mc\ in systems with giant 
or subgiant components (SG-Ch) has been underestimated. This is in fact a very likely possibility. 
A number of potential ways have been suggested to increase the frequency of SNe~Ia of this 
class (e.g.\ Hachisu, Kato \& Nomoto 1999). These ways include: (i)~mass stripping from the 
(sub-)giant companion by the strong wind from the white dwarf (this has the effect of increasing 
the range of mass ratios which result in stable mass transfer). (ii)~An efficient angular 
momentum removal by the stellar wind in wide systems (where the wind velocity and orbital 
velocity are comparable; this increases the range of  binary separations with result in 
interaction). While large uncertainties plague both of these suggestions (see Livio 1999 for 
a discussion), it is definitely possible that some physical processes which have not yet been 
properly included in the population synthesis calculations, will result in a significant 
increase in the rates from the channel with giant or subgiant companions. This means that 
in principle, the curve describing the SG-Ch channel (subgiant donor) in Fig.~7 (with the 
$z$-dependent dust opacity), may have to be shifted upwards (essentially parallel to itself, 
because of the involved delays). The curve could be shifted just enough for double-degenerates 
to dominate at redshifts $z\aplt0.5$, while SG-Ch dominate at $z\apgt0.5$. The question is 
now, could such a dominance shift be responsible for the apparent need for a cosmological 
constant? The answer is that this is definitely possible {\it in principle\/}. In particular, 
it has recently been suggested that the fiducial risetime of nearby SNe~Ia is $\sim2.5$~days 
longer than that of high-redshift SNe~Ia (Riess et~al.\ 1999a,b). It is far from clear though, 
whether such a change in the risetime (if real) could be attributed to different progenitor 
classes or to other evolutionary effects. One possibility could be that because SNe~Ia 
resulting from double-degenerates (if they indeed occur; Livio 1999) may have different 
surface compositions from those resulting from subgiant donors, this could affect the risetime. 
We would like to note, however, that we find the possibility of one progenitor class dominating 
at low redshifts and another at high redshifts rather {\it unlikely\/} (see also Livio 1999).   
The reason is very simple. As Fig.~7 shows, even if the SG~curve were to be shifted upwards, 
the result would be that the local (low~$z$) sample would have to contain a significant 
fraction of the SNe resulting from the SG~channel. Therefore, unless SNe~Ia from the SG~channel 
conspire to look identical to those from double-degenerates at low $z$, but different at high $z$, 
this would result in a {\it much less homogeneous\/} local sample than the observed one 
(which has 80--90\% of all SNe~Ia being nearly identical ``branch normals''; e.g.\ Branch 
1998 and references therein). Consequently, it appears that the observational indication 
of the existence of a cosmological constant cannot be the result of us being ``fooled'' 
by different progenitor classes (this does not exclude the possibility of other evolutionary effects). 

Finally, our models indicate that a careful determination of the rates of SNe~Ia as a function 
of redshift can place significant constraints on the cosmic star formation history, and on 
the significance of dust obscuration.

\acknowledgements

This work was supported by the Russian Foundation for Basic Research grant No.\ 96-02-16351. 
LRY acknowledges the hospitality of the Space Telescope Science Institute. ML acknowledges 
support from NASA Grant NAG5-6857. We acknowledge helpful discussions with N.~Chugaj, 
M.~Sazhin, and A.~Tutukov, and useful comments by D.~Branch.

\newpage 
\begin{figure}
\caption{Evolutionary scenarios for the most 
``productive'' potential progenitors of SNe~Ia. Ch---accumulation of a Chandrasekhar mass 
by a white dwarf and central carbon ignition. ELD---accumulation of 0.15 \msun\ of He on 
top of a subChandrasekhar mass white dwarf. See text for details.}
\end{figure}
\begin{figure}
\caption{The rates of potentially explosive events after an instantaneous star formation 
burst. The rates are normalized to a formation of 4.7 \msun\ of stars per yr. He-ELD---edge-lit 
detonations in systems with nondegenerate He donors; DD-Ch mergers of double degenerates 
with a total mass above \mc; SG-ELD---edge-lit detonations in systems  with subgiant donors; 
SG-Ch---accumulations of \mc\  in systems with subgiant donors. }
\end{figure}
\begin{figure}
\caption{Number counts of  SNe  per unit $\Delta z$\ {\it vs\/}.\ redshift for different 
channels of explosive events in Models~1 (``low dust'') and~2 (``$z$-dependent dust opacity''). 
Notations as in Fig.~2.}
\end{figure}
\begin{figure}
\caption{Cumulative counts of SNe below a given redshift $z$ for different  channels of 
explosive events in Models~1 and~2. Notations as in Fig.~2.}
\end{figure}
\begin{figure}
\caption{Number counts of  SNe per unit $\Delta z$\ {\it vs\/}.\ redshift for different 
channels of explosive events and different parameterized star formation histories 
(Models~3 to~6). Notations as in Fig.~2.}
\end{figure}
\begin{figure}
\caption{Cumulative counts of SNe below a given redshift $z$ for different channels of 
explosve events in Models~3 to 6.
Notations as in Fig.~2.}
\end{figure}
\begin{figure}
\caption{The ratios of the rates of different possible SNe channels to core-collapse SNe, 
for different assumptions on the dust and the SFR (see text). Notations as in Fig.~2. }
\end{figure}
\end{document}